\documentclass[aps,prb,twocolumn,superscriptaddress,showpacs]{revtex4}
\usepackage{amsmath,amssymb,bm}
\usepackage{graphicx,epsfig,verbatim}
\begin{document}

\title{Effects of Strain on Electronic Properties of Graphene}

\author{Seon-Myeong Choi}
\affiliation{Department of Physics, Pohang University of Science and Technology, Pohang 790-784, Korea}
\author{Seung-Hoon Jhi}
\email[Email:\ ]{jhish@postech.ac.kr}
\affiliation{Department of Physics, Pohang University of Science and Technology, Pohang 790-784, Korea}
\affiliation{Division of Advanced Materials Science, 
Pohang University of Science and Technology, Pohang 790-784, Korea}
\author{Young-Woo Son}
\email[Email:\ ]{hand@kias.re.kr}
\affiliation{Korea Institute for Advanced Study, Seoul 130-722, Korea}

\begin{abstract}
We present first-principles calculations of electronic
properties of graphene under uniaxial and isotropic strains, respectively.
The semi-metallic nature is shown to
persist up to a very large uniaxial strain of 30\%
except a very narrow strain range where a tiny energy gap opens.
As the uniaxial strain increases along a certain direction,
the Fermi velocity parallel to it decreases quickly
and vanishes eventually, whereas the Fermi velocity
perpendicular to it increases by as much as 25\%.
Thus, the low energy properties with small uniaxial strains
can be described by the generalized Weyl's equation
while massless and massive electrons coexist with large ones.
The work function is also predicted to increase
substantially as both the uniaxial and isotropic strain increases.
Hence, the homogeneous strain in graphene can be regarded
as the effective electronic scalar potential.
\end{abstract}
% 81.05.Uw Carbon, diamond, graphite
% N 62.25-g Mechanical properties of nanoscale systems
% 73.90+f Other topics in electronic stucture and electrical properties of surface,,...low dim materials.

\maketitle
Mechanical strain often gives rise to surprising effects
on electronic properties of carbon
nanomaterials~\cite{heyd,yang,tombler,han,minot}.
It can turn the metallic nanotube into semiconductor
and vice versa~\cite{heyd,yang,tombler,han,minot}.
Along with the uniquely strong mechanical properties
of the $sp^2$- and $sp^3$- bonded carbon materials~\cite{ruoff},
the interplays between mechanical and electronic properties
may be useful in various applications~\cite{sazonova}.
A recent successful isolation of a new
carbon allotrope~\cite{novoselov1}, graphene,
offers a new opportunity to explore
such interesting electromechanical properties
in two dimensions.

%Graphene is a single layer of carbon atoms in
%the hexagonal lattice having two sublattices~\cite{rmp}.
At low energies, graphene at equilibrium has
two linear energy bands that intersect each other
at the high symmetric points, $K$ and $K'$,
of the first Brillouin zone (BZ) and are isotropic
with respect to the points~\cite{rmp}.
Without strains,
the density of states vanishes linearly at the Fermi energy ($E_F$)
or the Dirac point ($E_D$), exhibiting a semi-metallic nature.
Thus, charge carriers are well described by the Dirac's equation
for a (2+1)D free massless fermion~\cite{rmp,novoselov2,pkim}.
Electron states here have another quantum number
called a pseudospin which is either parallel or antiparallel
to the wavevector of the electron and is of central
importance to various
novel phenomena~\cite{rmp,novoselov2,pkim,chpark1,chpark2}.
Mechanical strains can introduce new environments in studying such
novel physics of graphene.

Recently, several experiments have been performed to investigate
the physical properties of graphene when its hexagonal lattice
is stretched out of
equilibrium~\cite{ni,hong,ferrari,hone1,changgu,teague,ferralis}.
Strain can be induced on graphene either intentionally or naturally.
The uniaxial strain can be induced by bending
the substrates on which graphene is elongated without slippage~\cite{ni,hong,ferrari,hone1}.
Elastic responses are measured
by pushing a tip of atomic force microscopes on suspended graphene~\cite{changgu}.
Graphene on top of SiO$_2$~\cite{teague} or SiC surface~\cite{ferralis}
also experiences a moderate strain due
to surface corrugations or lattice mismatch.
Motivated by recent works~\cite{ni, khare, liu, antonio_strain,farjam}
pointing to a remarkable stability of graphene
with large strains, we have carried out first-principles calculations
and theoretical analysis to explore
the electronic structures of strained graphene and to understand
its low energy electronic properties.

In this Rapid Communication, we show that no sizable energy gap
opens in uniaxially strained graphene and the variation in energy bands
strongly depends on the direction of uniaxial strains.
We also predict that the work function increases substantially
as both the uniaxial and isotropic strain increases.
When an uniaxial strain less than 26.2\% is applied
along the zigzag chain direction,
the semi-metallicity is sustained.
Beyond that, the system develops a small energy gap up to 45.5 meV
at a strain of 26.5\% and then close
its gap quickly due to the downshift of the $\sigma^*$ band to the $E_F$.
This differs from conclusions of the previous literatures~\cite{ni, liu, antonio_strain}.
With uniaxial strain along the armchair chain direction,
no energy gap develops.
Under uniaxial strain,
the group velocities at the $E_F$ are shown to be strong functions of the wavevectors
so that the low energy properties
with small uniaxial strains can be described by the generalized
Weyl's equation~\cite{chpark1,chpark2,organic,hasegawa,koba,tajima}.
With large uniaxial strains, quasiparticles become massive
along the strain direction while ones in the perpendicular direction are still massless.
%With the isotropic strain, no energy gap opens and the group velocity
%decreases linearly with increasing the strain.

Computations were carried out using the pseudopotential density functional method
with a plane-wave basis set~\cite{kresse}.
%implemented in the Vienna {\it ab initio} software package~\cite{kresse}.
The exchange-correlation interactions were treated within
the Perdew-Berke-Enzelhof~\cite{pbe} generalized gradient approximation.
The cutoff energy for expansion of wave-functions and potentials was 400 eV
and the Monkhorst-Pack k-point grid of $12\times12\times1$ is used
for the atomic relaxation and of $60\times60\times1$ for electronic structure calculations.
The atomic relaxation was carried out until the change
in the total energy per one unit cell was smaller than 0.1 meV.
The layer-to-layer distance between adjacent graphene in the supercells is 15.0 \AA.
%avoiding spurious interactions between them.

\begin{figure}[t]
\centering
\includegraphics[width=0.85\columnwidth]{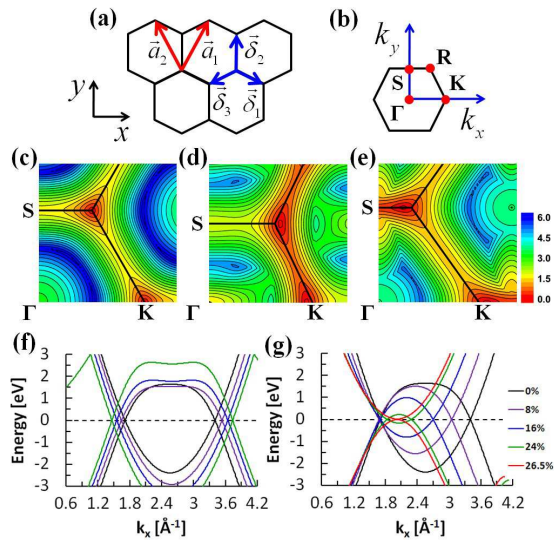}
\caption
{(color online) (a) Hexagonal lattice of graphene.
${\bm a}_1$ and ${\bm a}_2$ are the lattice vectors.
With $Z(A)$-strain,
${\bm a}_1=(a_x, a_y)$ and  ${\bm a}_2=(-a_x, a_y)$.
${\bm\delta}_i$ ($i=1,2,3$) connects three nearest neighbors.
(b) The first BZ with high symmetric points.
Energy contours for graphene (c) without strain,
(d) $A$-strain of 20\%
and (e) $Z$-strain of 20\%.
The scale bar for contours is in unit of eV.
The $\pi$ and $\pi^*$ bands with various
(f) $A$-strains and (g) $Z$-strains along the line
of $k_y=0$ in (b).
}
\end{figure}

Here, we consider graphene only under uniaxial and isotropic strains, respectively.
For comparison, the electronic structures of graphene
under uniaxial strains along the two special directions are investigated.
The effects of uniaxial strain along arbitrary directions
and those of isotropic strains will also be discussed later.
Following previous conventions~\cite{antonio_strain},
the uniaxial strain along the zig-zag chain direction [$x$-axis  in Fig. 1(a)]
in the honeycomb lattice is denoted by the $Z$-strain
and one perpendicular to this ($y$-axis) by the $A$-strain.
%The effects of isotropic strain
%will be discussed in the final part of the paper.
From the fully relaxed atomic geometries,
the calculated Poisson's ratios
for graphene as functions of the magnitude and direction of strains
agree with the previous calculations~\cite{liu,farjam}.

We find that if the magnitude of strain is less than 26.2\%,
no gap opens with the $Z$-strain. Graphene with the $A$-strain
also has no energy gap up to a magnitude of 30\%.
As shown in the energy contour from first-principles calculations,
the $E_D$'s coincide with the high symmetric
$K$ and $K'$ (or $R$) points of the first BZ without strains (Fig. 1 (c)).
With the $A$-strain, the $E_D$'s are off the symmetric points and
the two adjacent $E_D$'s along the $k_y=0$ line
repel each other as the strain increases (Figs. 1(d) and (f)),
agreeing with previous calculations~\cite{antonio_strain}.
Contrary to the cases with the $A$-strain,
the two adjacent $E_D$'s with the $Z$-strain approach each other
(Figs. 1(e) and (g)) and merge together eventually
at strain of 26.2\%.
%It is also noticeable that the energy differences between conduction
%and valence bands at the $S$-point increase as the $A$-strain increases
%while ones decrease as the $Z$-strain increases (Figs. 1(f) and (g)).

The mismatch of the Dirac points with the high symmetric BZ points
can be easily understood by one-orbital tight-binding
approximations~\cite{antonio_strain,organic,hasegawa}.
In the elastic regime under the $Z$-strain,
the kinetic hopping integrals ($t$) between
the nearest neighbors will depend on its
connecting vectors, ${\bm\delta}_i$ ($i=1,2,3$)
such that $t_1 = t_3 < t_2$ where
$t_i\equiv t({\bm\delta}_i)$ ($i=1,2,3$) (Fig. 1(a)).
Under the $A$-strain, $t_1=t_3>t_2$.
Considering the nearest-neighbor hoppings only,
the Hamiltonian of graphene with $Z$($A$)-strain
can be written as
${\mathcal H}=-t_2 \sum_{\bm k}\left[\xi({\bm k})c^\dagger_{A{\bm k}}c_{B{\bm k}}
+c.c.\right]$
where $\xi({\bm k})=e^{\bm{k}\cdot\bm{\delta}_2}(1+2\eta e^{-ik_y a_y}\cos(k_x a_x))$,
$\eta\equiv t_1 /t_2=t_3/t_2$, ${\bm k}=(k_x, k_y)$ and
$c_{A(B){\bm k}}$ is an annihilation operator for an electron
with momentum $\bm k$ on the sublattice $A(B)$.
The resulting energy dispersion is given by $E_{\bm k}=\pm t_2|\xi({\bm k})|$.
$\eta<1$ $(\eta>1)$ for the $Z(A)$-strain.
On the $k_y=0$ line in the first BZ, the $x$-component of $K$-point is given by
$k_K=\frac{\pi}{2a_x}\left(1+\frac{a_x^2}{a_y^2}\right)$ whereas the Dirac point with strains,
i.e., the zero energy solution, $\xi(k_D,0)=0$,
is given by $k_D=\frac{1}{a_x}\cos^{-1}(-\frac{1}{2\eta})$.
Hence, under the $A(Z)$-strain, $k_D \neq k_K$ as shown
in Fig. 1.

We find that the energy splitting between the $\sigma$ and $\sigma^*$ bands
at the $S$ point is reduced when the $Z$-strain increases (Fig. 2)
and one at the $\Gamma$ point does with the $A$-strain (not shown here).
The strain-induced small energy gap is eventually closed due to downshift
of the $\sigma^*$ band at the $Z$-strain of 27\% (Fig. 2).
In very high strain regime, a single orbital tight-binding
approximation fails to capture the downshift of $\sigma^*$-orbitals
although it shows approximately similar variations of
$\pi$-bands in the low and moderate strain regimes~\cite{antonio_strain}.
It is noticeable that $\pi$ ($\pi^*$) electrons along $SR$ become massive
but that those along $S\Gamma$ are still massless
after the gap closure (Fig. 2).
Anomalous area expansion, i.e., the negative Poisson's ratio~\cite{lakes} is found
when the $\sigma^*$ band touches the $E_F$ at the $Z$-strain larger than 27\%
because the antibonding states are occupied (the unit cell area increases
by 35\% under the $Z$-strain of 30\%).
However, at this point, graphene may not be stable~\cite{hong,liu}.
Hereafter, we will consider graphene with strains less than 26.5\%~\cite{liu}.

\begin{figure}[t]
\centering
\includegraphics[width=0.9\columnwidth]{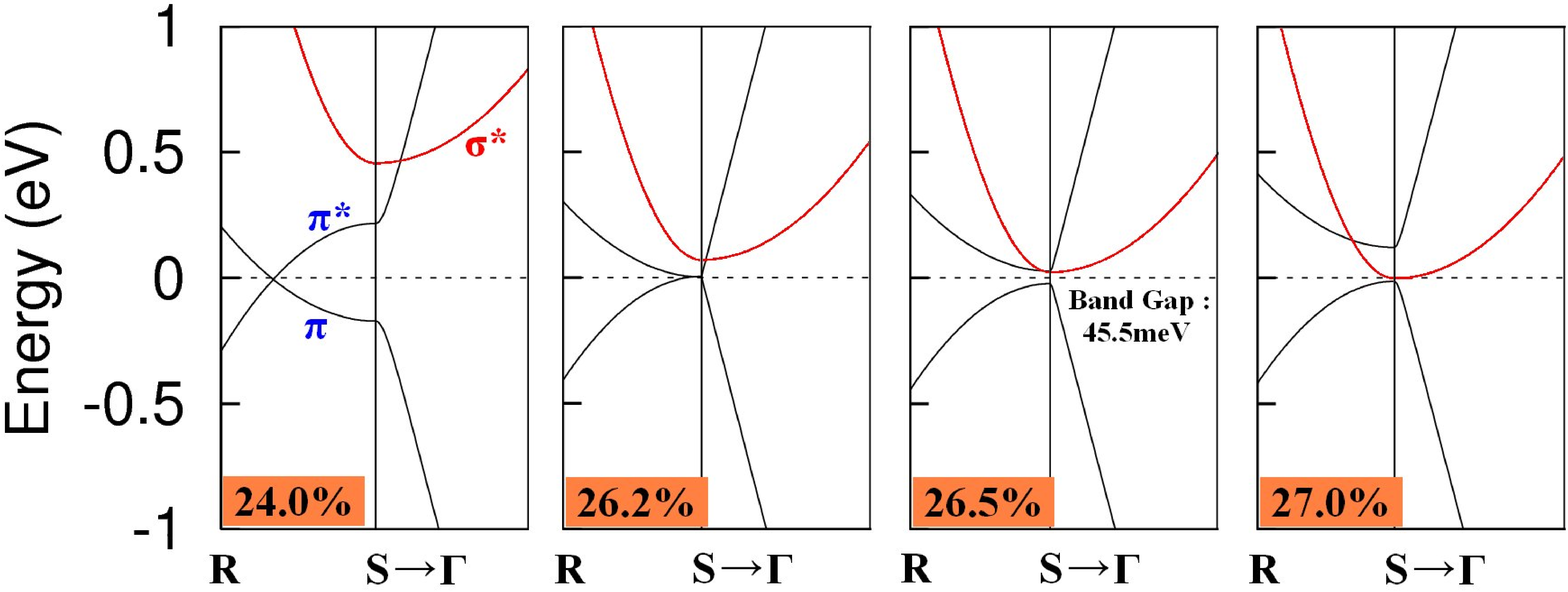}
\caption
{(color online) Calculated band structures of the strained graphene
around the $S$ point with a $Z$-strain
of 24.0\%, 26.2\%, 26.5\% and 27.0\% (from left to right panels),
respectively.}
\end{figure}

As uniaxial strain increases, the group velocity
at the $E_D$ increases or decreases substantially
depending on the wavevectors (Fig. 3).
We calculate the group velocities of electrons by differentiating
the energy dispersion of conduction bands directly, i.e.,
$v_{\bf k}=\frac{1}{\hbar}
\left[\frac{\partial E_{\bf k}}{\partial {\bf k}}\right]_{E_{\bf k}=E_F}$.
The group velocity along the $A$-strain ($v_{A3}$ in Fig. 3)
decreases as increasing strain
while ones ($v_{A1}$ and $v_{A4}$) in direction perpendicular to strains increase.
Up to the $A$-strain of 24\%, $v_{A3}$ is reduced by almost 60\%
of the group velocity without strains ($v_0$) and
 $v_{A1}$ and $v_{A4}$ increase linearly by 25\%.
We also find that $v_{A1}$ differs $v_{A4}$ (opposite direction to the former)
as shown in Fig. 3(a).
Along the specific direction 2 in insets of Fig. 3,
$v_{A2}\simeq v_{Z2}\simeq v_0$.
Under the $Z$-strain, the similar behaviors occur (Fig. 3(b)).
It is also noticeable that $v_{Z1}$ differs $v_{Z4}$
(opposite direction to the former) and the both become zero
at the strain of 26.2\%.
We note that the group velocity anisotropy under strains
may lead to an anisotropy of resistance
shown in a recent experiment~\cite{hong}.

The low energy properties of graphene with moderate strains as revealed by
our first-principles calculatioins can be described well
by the generalized Weyl's equation~\cite{organic,koba,tajima}.
By expanding $\xi({\bf k})$ around ($k_D$,0)
up to the first order of small momentum $\bf q$,
$\xi({\bf q})=\xi(k_D+q_x,q_y)\simeq
(4\eta^2 -1)^{1/2}a_x q_x-ia_y q_y$.
The resulting Hamiltonian can be written as
${\mathcal H}\simeq v_x \sigma_x q_x + v_y \sigma_y q_y$ where
$\sigma_{x(y)}$ are Pauli matrices, $v_x=t_2 a_x (4\eta^2 -1)^{1/2}$
and $v_y=t_2 a_y$.
With the $Z$-strain, $t_2$ increases predominantly
over a contraction of $a_y$~\cite{porezag} so that $v_y$ increases.
On the other hand,
$v_x=t_2 a_x (4\eta^2 -1)^{1/2}=t_1 a_x (4-1/\eta^2)^{1/2}<\sqrt{3}t_1 a_x$
since $\eta<1$ with the $Z$-strain.
Hence, $v_x$ decreases very quickly upon elongation of $a_x$
followed by reduction of $t_1$ with the $Z$-strain.
For the $A$-strain, the opposite situation occurs.
We note that this Hamiltonian also describes the low energy physics of
graphene superlattice~\cite{chpark1,chpark2}
and $\alpha$-(BEDT-TTF)$_2$I$_3$~\cite{organic,koba,tajima} respectively.
Thus, like graphene superlattices~\cite{chpark2},
the pseudospin in uniaxially strained graphene is not in parallel
or antiparallel to the wavevectors suggesting
some interesting transport properties~\cite{chpark1,antonio_strain2,fogler}.

\begin{figure}[t]
\centering
\includegraphics[width=0.9\columnwidth]{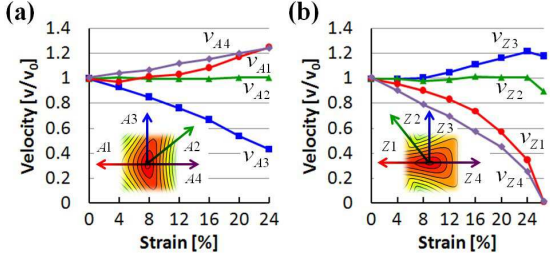}
\caption
{(color online) The group velocities ($v_{Ai}$) of $\pi$ and $\pi^*$ electrons with the $A$-strain (a)
and $v_{Zi}$ with the $Z$-strain along the direction $i$ ($=1,2,3,4$ in insets)
in an unit of isotropic group velocity ($v_0$) without strain.
The angle between the direction A2 and A3 in inset of (a) is 52$^\circ$
and one between Z2 and Z3 in (b) is 38$^\circ$.
}
\end{figure}

Although the simple model described above can explain the results from our
first-principles calculations in general, we should point out that
the next-nearest neighbor (nnn) hopping, $t'$, plays an important role
in the low energy properties~\cite{organic}.
As shown in Fig. 3,
the group velocities along $+\hat{k}_x$ ($v_{A(Z)4}$)
differ one along $-\hat{k}_x$ ($v_{A(Z)1}$)
implying tilted anisotropic Dirac cones due
to the nnn interactions~\cite{organic,koba,tajima}.
With the $A(Z)$-strain, $t'$ also depends on its six connecting vectors,
such that $t'_\alpha\equiv t'(\pm{\bf a}_1) = t'(\pm{\bf a}_2)$ and
$t'_\beta\equiv t'(\pm({\bf a}_1-{\bf a}_2))$.
$\chi\equiv t'_\beta /t'_\alpha < 1 (>1) $ for the $Z(A)$-strain.
The effective Hamiltonian for the nnn interactions around
$(k_D,0)$ can be written as
${\mathcal H}'\simeq v'_x q_x \sigma_0$ where
$v'_x=a_x t'_\alpha\left(1-\chi/\eta\right)(4-1/\eta^2)^{1/2}$
and $\sigma_0$ is an identity.
The resulting energy dispersion can be expressed concisely
as ${\mathcal E}_{\bf q}=v(\phi_{\bf q})q$
where $\phi_{\bf q}=\tan^{-1}(q_y /q_x)$, $q=(q^2_x+q^2_y)^{1/2}$
and
$v(\phi_{\bf q})=v'_x\cos\phi_{\bf q}
\pm(v^2_x \cos^2\phi_{\bf q}+v^2_y\sin^2\phi_{\bf q})^{1/2}$~\cite{organic,koba,tajima}.
So, $v_{A1(4)}=v_x\mp v'_x$ and $v_{Z1(4)}=v_x\pm v'_x$ as shown in Fig. 3.
Hence, the Dirac cone is tilted in the $k_x$ direction
regardless of the uniaxial strain direction.
With the large strain ($>$ 20\%), $v'_x$ becomes negligible so that
$v_{A(Z)1}\simeq v_{A(Z)4}$.
And, the tilting effect disappears when the large uniaxial strain is applied.

The density of states ($D(E)$) around the $E_D$ increases
gradually as the uniaxial strain increase while maintaining its linearity (Fig. 4).
With large strains ($>$ 20 \%), $D(E)$ shows an abrupt change
depending on the direction of strains.
Figure 4 shows the calculated $D(E)$ with $|E-E_F|<0.6$ eV
from first-principles calculations.
From the generalized Weyl's equation,
$D(E)=\frac{2}{\pi}\frac{|E|}{\bar{v}^2_F}$
where $1/\bar{v}^{2}_F=\frac{1}{2\pi}
\int^{2\pi}_{0}d\phi_{\bf q}/v^{2}(\phi_{\bf q})$~\cite{organic,koba,tajima}.
The strain-induced reductions
in the averaged anisotropic group velocities ($\bar{v}^{2}_F$)
will increase the slope of $D(E)$ as shown in Fig. 4.
The $D(E)$ changes significantly when $\sigma^*$ band is near the $E_F$
with the Z-strain (Fig. 4(b)).
With the large Z-strain, the merging of two Dirac points signals
the van Hove singularities of the $\pi$ and $\pi^*$ bands
(24\% case in Fig. 4(b)).
When the gap opens with the $Z$-strain of 26.5\%,
$D(E<E_F)\sim\sqrt{E}$~\cite{hasegawa} and $D(E>E_F)$ shows a steep enhancement due to
the $\sigma^*$ band.

\begin{figure}[t]
\centering
\includegraphics[width=1.0\columnwidth]{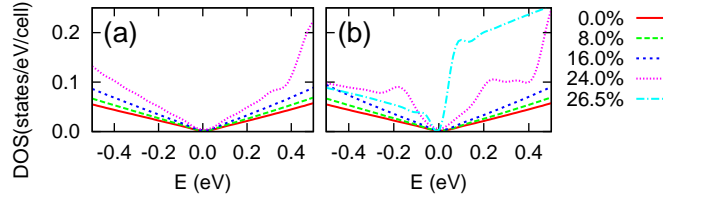}
\caption
{(color online) Calculated density of states of graphene with (a) the $A$-strain and (b) $Z$-strain.}
\end{figure}

\begin{figure}[b]
\centering
\includegraphics[width=0.6\columnwidth]{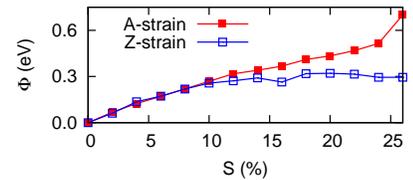}
\caption
{(color online) Calculated work functions ($\Phi$) of graphene with the $A$- and $Z$-strain.}
\end{figure}

The work function in uniaxially strained graphene is predicted
to increase substantially as the strain increases (Fig. 5).
The calculated work function of graphene without strain is
4.5 eV agreeing with the previous theoretical~\cite{kelly}
and experimental~\cite{oshima} estimations.
As the strain increases up to 12\%, the work function increases linearly
by 0.3 eV regardless of the direction of strains as shown in Fig. 5.
The work function rises up further to 5.2 eV as the $A$-strain reaches 26\%.
However, with larger $Z$-strains, the work function saturates to 4.8 eV.
Hence the variations in the work function can also characterize
the direction of the strain.
Our calculated results indicate that the controlled charge transfer
between gaseous molecules and graphene can be realized by straining graphene.
We also anticipate that the strain affects the band lineup
at the graphene-metal contact~\cite{kelly}.

To study the effect of uniaxial strains in arbitrary directions,
we study the band structure of graphene stretched
along the direction rotated by 10.9$^\circ$ with respect to the
$x$-axis in Fig. 1(a).
We confirm that no energy gap opens
up to a strain of 30\% (not shown here).
The work function also increases as strain increases.
Our {\it ab initio} calculations
conclude that no energy gap opens under uniaxial strain less than 26\%
along any arbitrary direction.

Finally, we calculate the variations of electronic properties of graphene
under the isotropic strain ($I$-strain). Because the $I$-strain maintains all
crystal symmetries of graphene,
the electronic structures show no significant changes unlike uniaxially strained
cases. The Fermi velocity decreases linearly to 86\% of $v_0$
as the $I$-strain increases up to 10\% (Fig. 6 (a)).
The work function of the system also increases linearly up to 0.64 eV
as the $I$-strain reaches 10\% (Fig. 6(b)).
From the calculation results,
it is shown that the uniform strain induces
effective vector~\cite{rmp} and electric scalar potential in graphene.

In summary, from first-principles calculations,
it is shown that strained graphene does not develop an energy gap
and that the group velocities under uniaxial strain exhibit
a strong anisotropy.
%The next-nearest neighbor interactions are found to be
%crucial to describe the low energy properties of graphene with uniaxial strains.
We show that the generalized Weyl's equation
is an appropriate model for uniaxially strained graphene
that incorporates all assessed properties that go beyond the simple
tight-binding approximations.
It is also shown that the work function of strained graphene 
increases substantially as strain increases.

\begin{figure}[t]
\centering
\includegraphics[width=1.0\columnwidth]{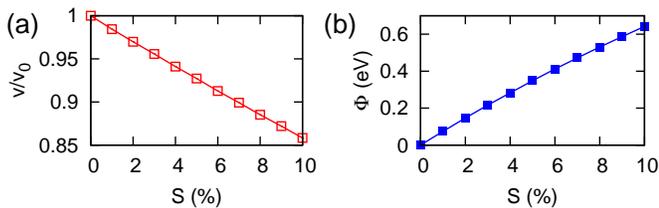}
\caption
{(color online) (a) Calculated Fermi velocity variation under the $I$-strain in an unit
        of the Fermi velocity ($v_0$) without strain. (b) Calculated
work functions ($\Phi$) the $I$-strain. We set the work function
without strain to zero here.}
\end{figure}

{\it Note added.}-- After submission, we became aware of related work 
on similar systems from other groups \cite{mohr,pucci}.

Y.-W. S. thank P. Kim, C.-H. Park, S. G. Louie, and C. Park
for valuable discussions.
S.-M. C. and S.-H. J. were supported by the KOSEF grant
funded by the MEST 
(No. R01-2008-000-20020-0 and WCU program No. R31-2008-000-10059-0) and
Y.-W. S. by Nano R\&D program, No.2008-03670 and 
Quantum Metamaterials research center, No. R11-2008-053-01002-0.

\end{document}